\documentclass[prb,twocolumn,showpacs,preprintnumbers,amsmath,amssymb,superscriptaddress]{revtex4}

\usepackage{graphicx}
\usepackage{dcolumn}
\usepackage{bm}
\usepackage{color}
\usepackage{adjustbox}

\begin{document}

\title{Sr$_3$Ir$_2$O$_7$F$_2$: Topochemical conversion of a relativistic Mott state into a spin-orbit driven band insulator}

\author{Christi Peterson}
\altaffiliation{Contributed equally to this work}
\author{Michael W. Swift}
\altaffiliation{Contributed equally to this work}
\author{Zach Porter}
\altaffiliation{Contributed equally to this work}
\author{Rapha{\"e}le J. Cl{\'e}ment}
\affiliation{Materials Department, University of California, Santa Barbara, California 93106-5050, USA}
\author{Guang Wu}
\affiliation{Department of Chemistry and Biochemistry, University of California, Santa Barbara, California 93106-5050, USA}
\author{G. H. Ahn }
\affiliation{Department of Physics, Hanyang University, Seoul 04763, Korea}
\author{S. J. Moon}
\affiliation{Department of Physics, Hanyang University, Seoul 04763, Korea}
\author{B. C. Chakoumakos}
\affiliation{Neutron Scattering Division, Oak Ridge National Laboratory, Oak Ridge, Tennessee 37831, USA}
\author{Jacob P. C. Ruff}
\affiliation{Cornell High Energy Synchrotron Source, Cornell University, Ithaca, New York 14853, USA}
\author{Huibo Cao}
\affiliation{Neutron Scattering Division, Oak Ridge National Laboratory, Oak Ridge, Tennessee 37831, USA}
\author{Chris Van de Walle}
\author{Stephen D. Wilson}
\email{stephendwilson@ucsb.edu}
\affiliation{Materials Department, University of California, Santa Barbara, California 93106-5050, USA}

\begin{abstract}

The topochemical transformation of single crystals of Sr$_3$Ir$_2$O$_7$ into Sr$_3$Ir$_2$O$_7$F$_2$ is reported via fluorine insertion.  Characterization of the newly formed Sr$_3$Ir$_2$O$_7$F$_2$ phase shows a nearly complete oxidation of Ir$^{4+}$ cations into Ir$^{5+}$ that in turn drives the system from an antiferromagnetic Mott insulator with a half-filled J$_{eff}=1/2$ band into a nonmagnetic $J=0$ band insulator.  First principles calculations reveal a remarkably flat insertion energy that locally drives the fluorination process to completion. Band structure calculations support the formation of a band insulator whose charge gap relies on the strong spin-orbit coupling inherent to the Ir metal ions of this compound.   

\end{abstract}

\maketitle

\section{Introduction}
Topochemical transformations have long been harnessed to control the properties of a broad array of quantum materials. Post growth transformations of bulk single crystals are often utilized to achieve a number of unconventional electronic states ranging from quantum spin liquids \cite{abramchuk2017cu2iro3} to charge density waves \cite{Zhang8945} to superconductivity \cite{delville}.  The power of this approach is its utility in modifying the properties of an existing crystal framework to realize ground states and properties \cite{bocarsly2013superconducting, zhang2016directed} otherwise precluded from conventional synthetic methods \cite{mccabe2007fluorine}. 

One area of interest is the potential of exploring new electronic states in spin-orbit assisted Mott materials.  These are systems where the interplay between strong crystal field energies,  spin-orbit coupling, and on-site Coulomb interactions cooperate to stabilize a Mott insulating state comprised of spin-orbit entangled electrons \cite{kim2008novel}. Control over carrier concentrations \cite{hogan2015first} and dimensionality/bandwidth \cite{PhysRevLett.101.226402} in these systems are key to realizing many of the new states predicted \cite{wang2011twisted, jackeli2009mott, wan2011topological}; however these degrees of freedom are often constrained by the solubility limits of conventional dopants and the structural stabilities of the host lattices \cite{PhysRevB.94.195115}.  

Modifying the host lattices of these unusual spin-orbit Mott insulators via topochemical transformations has recently shown promise. Chemical reactions driving postgrowth transformations have stabilized spin liquid formation\cite{abramchuk2017cu2iro3, kitagawa2018spin}, unconventional metallic phases,\cite{takayama2014spin} as well as new magnetic phases \cite{roudebush2016iridium}. However, utilizing similar techniques to control the Ir valence and lattice dimensionality of the $n=1$ and $n=2$ members of the Ruddlesden-Popper (R.P.) series of strontium iridates Sr$_{n+1}$Ir$_n$O$_{3n+1}$, the seminal examples of spin-orbit assisted Mott states, remains notably unexplored. 

The development of novel routes of controlling the electronic states and lattice architectures of R.P. strontium iridates is of interest for a variety of reasons. For instance, doping holes into Sr$_{n+1}$Ir$_n$O$_{3n+1}$ is a proposed route to realizing unconventional superconductivity\cite{PhysRevB.89.094518} and controlling the relative spacing between IrO$_6$ planes is a potential mechanism for controlling their unusual magnetic ground states. \cite{PhysRevLett.114.247209} Trivial alloying of these systems to explore these effects has proven severely limited by solubility bounds \cite{PhysRevB.92.075125} as well as disorder effects associated with perturbing the IrO$_6$ planes \cite{clancy2014dilute, dhital2014carrier}.  To this end, an appealing alternate approach of controlling the dimensionality and valence of Sr$_{n+1}$Ir$_n$O$_{3n+1}$ compounds is through postgrowth reaction with fluorine gas.  Prior studies of bilayer ($n=2$) $3d$ and $4d$ transition metal oxide R.P. phases \cite{mccabe2007fluorine} demonstrate a common instability of this structure to the incorporation of F$^{-}$ anions and suggest that similar techniques may be leveraged within their $5d$ cousins.  

Here we report the fluorine-driven transformation of Sr$_3$Ir$_2$O$_7$ crystals into Sr$_3$Ir$_2$O$_7$F$_2$, which in turn raises the valence of Ir ions in the lattice from Ir$^{4+}$ to Ir$^{5+}$ and drives the system from an antiferromagnetic $J_{eff}=1/2$ Mott state into a nonmagnetic $J=0$ spin-orbit driven band insulator.  Soft postgrowth reaction of Sr$_3$Ir$_2$O$_7$ crystals in a low temperature fluorine environment drives a rapid expansion of the unit cell's long-axis (by $\approx 16\%$) as F$^{-}$ ions intercalate within the SrO rocksalt layers of the structure.  The reaction diffuses to completion (incorporating two F atoms per formula unit) throughout the bulk of the crystal due to a low insertion energy and shallow chemical potential gradient with increasing F content.  Experimental data combined with first principles calculations characterize the resulting Sr$_3$Ir$_2$O$_7$F$_2$ state as a highly distorted spin-orbit driven insulator, and our combined results demonstrate Sr$_3$Ir$_2$O$_7$F$_x$ as a unique material that can be driven from a $J_{eff}=1/2$ into a $J=0$ spin-orbit entangled ground state.   

\section{Experimental Details}
\subsection{Crystal growth}
Single crystals of Sr$_3$Ir$_2$O$_7$ were grown via a halide flux growth technique similar to earlier reports \cite{subramanian1994single, hogan2015first}.  Starting powders of SrCO$_3$, IrO$_2$, and SrCl$_2$ were mixed in a 1:1:15 molar ratio and placed inside of a platinum crucible with a lid.  The mixture was then reacted at 1300 $^{\circ}$C for 5 hours and slowly cooled to 850 $^{\circ}$C over 125 hours. Plate-like single crystals (typical size $1\times1\times0.3$ mm) were then removed from the flux by etching the resulting boule with deionized water.  

\subsection{Fluorination of Sr$_3$Ir$_2$O$_7$}
Once single crystals were extracted from the boule, their phase purity was verified with x-ray diffraction measurements, and crystals were then loaded into an alumina boat and buried within CuF$_2$ powder.  The crucible was placed inside of a tube furnace under argon gas flow and heated to 250 $^{\circ}$C over the course of 30 minutes.  This temperature was then held for 2 hours and the reaction was stopped by quenching the crucible in air. For reaction times substantially less than this, crystals were found to be mixed phase containing both the Sr$_3$Ir$_2$O$_7$F$_2$ and Sr$_3$Ir$_2$O$_7$ structures, and for reaction times appreciably greater than this, samples began to show local decomposition into SrF$_2$.  We note here that similar fluorination could be achieved via reaction in sealed quartz ampoules containing XeF$_2$, indicating that the F insertion process is driven via exposure to F$_2$ gas and not a more complicated process of anion exchange involving direct contact with the CuF$_2$.

\subsection{Neutron and x-ray diffraction measurements}
Neutron diffraction measurements were performed on the HB3A diffractometer at the High Flux Isotope Reactor at Oak Ridge National Laboratory. A Si-220 monochromator ($1.546$ \AA) was utilized and a 2D Anger camera detector collected the scattering intensity. All data were corrected for neutron absorption through WinGX \cite{farrugia2012wingx} and refined via the FullProf software package \cite{rodriguez1990fullprof}.  The minimum neutron transmission was 0.7085 for the (0 0 4) reflection and the maximum neutron transmission was 0.82662 for (2 0 0) reflection with an average transmission of 0.78414. X-ray powder diffraction data were collected on powder samples comprised of crushed crystals and were collected on an Panalytical Empyrean Powder Diffractometer with a Cu source.  Single crystal x-ray data were collected on a Kappa Apex II Diffractometer with a Mo source ($\lambda=0.71073$ \AA).

\subsection{Charge transport/magnetization}
Charge transport measurements were performed in a Quantum Design Physical Property Measurement System (DynaCool PPMS) with an AC resistance bridge. Crystals were mounted in a four wire configuration and silver paint used to create contacts.  Transport data were collected with current flowing within the $bc$-plane. Magnetization measurements were performed within a SQUID-based Quantum Design Magnetic Property Measurement System (MPMS3), and crystals were mounted on a quartz paddle with a small drop of varnish.  Magnetization data were collected with the field applied within the $bc$-plane. 

\begin{figure}[t]
	\includegraphics[scale=.45]{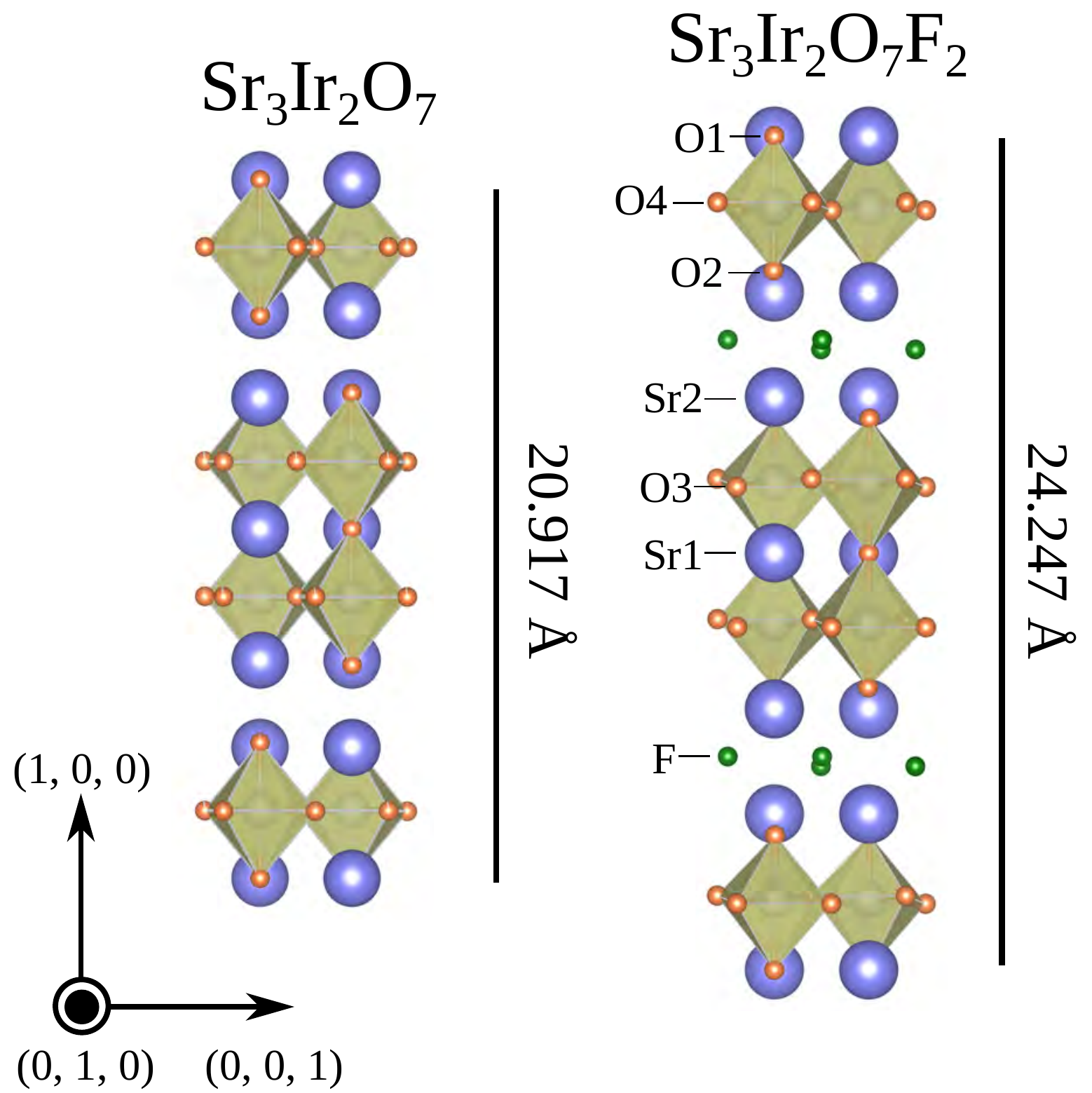}
	\caption{Illustration of the crystallographic unit cells of Sr$_3$Ir$_2$O$_7$ and Sr$_3$Ir$_2$O$_7$F$_2$.  IrO$_6$ octahedra are shaded and purple spheres denote Sr atoms.  Green spheres denote the positions of intercalated F$^{-}$ anions.  The long axis of the parent monoclinic unit cell expands by 16$\%$ upon incorporation of two F atoms per formula unit into the crystal matrix.}
\end{figure}

\begin{figure}[h]
	\includegraphics[scale=.65]{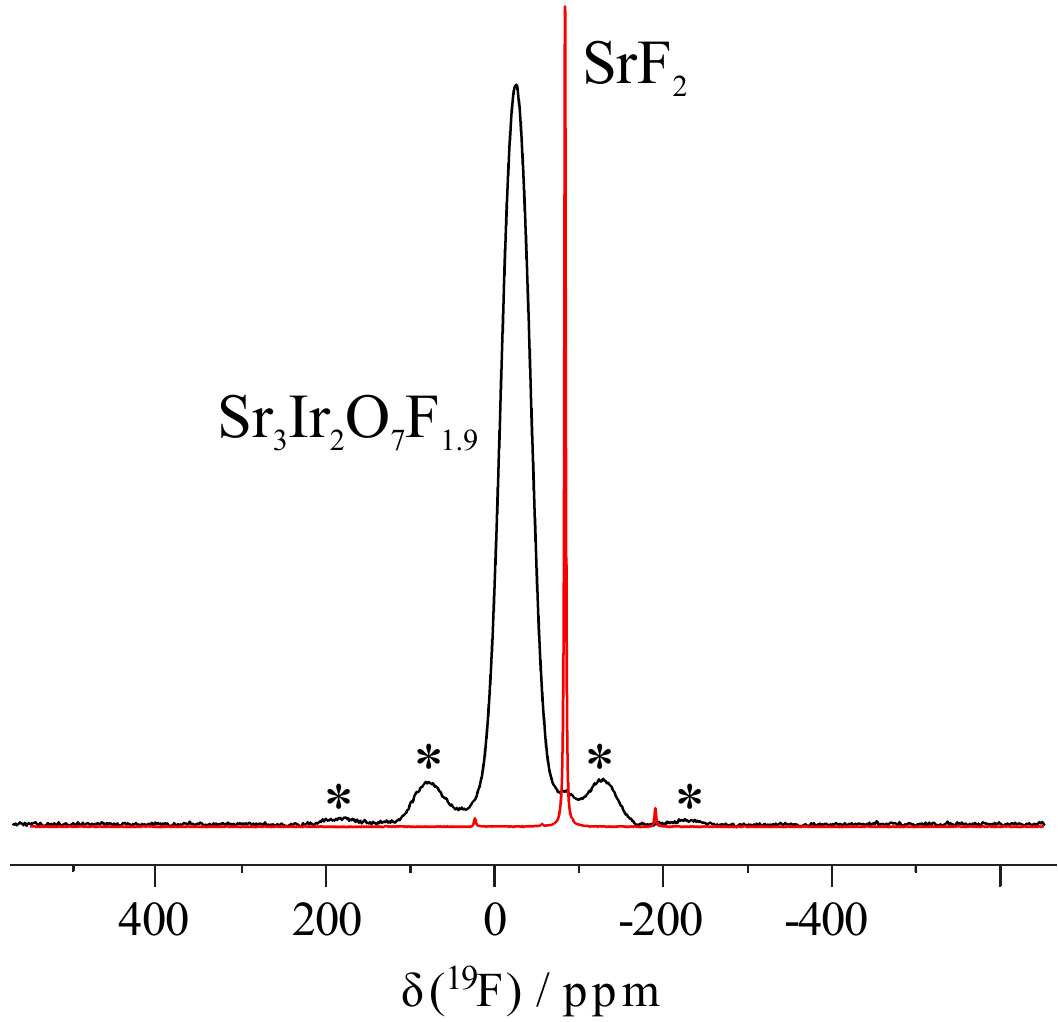}
	\caption{$^{19}$F spin echo NMR spectra illustrating a dominant F environment with a chemical shift of $\delta=25.1$ ppm in a measurement of crushed Sr$_3$Ir$_2$O$_7$F$_2$ crystals.  Asterisks denote spinning sidebands associated with this peak.  A weak secondary signal is observed at $\delta=88$ ppm beneath the centerband and the sideband of the primary site, which matches the shift expected for a SrF$_2$-like F environment.  The black line denotes data for Sr$_3$Ir$_2$O$_7$F$_2$ and the red line denotes data for the reference SrF$_2$.}
\end{figure}

\subsection{Optical conductivity measurements} 
Optical reflectivity measurements were also performed to characterize the electronic structure of Sr$_3$Ir$_2$O$_7$F$_2$.  The $bc$-plane reflectivity spectra $R(\omega)$ was measured in the energy region between 5 meV and 1 eV by using an in-situ overcoating technique \cite{homes1993cc}. The complex dielectric constant between 0.74 eV and 5 eV was obtained by using a spectroscopic ellipsometer, and the complex optical conductivity $\sigma(\omega)=\sigma_1(\omega)+i\sigma_2(\omega)$ was determined via a Kramers-Kronig analysis\cite{wooten}.

\subsection{X-ray Absorption Measurements}
X-ray absorption spectroscopy (XAS) measurements were performed at Beamline A2 at CHESS. Crushed single crystals were prepared on three layers of Kapton tape, with an average sample thickness corresponding to nearly two absorption lengths. Measurements were collected at the Ir L$_2$ ($2p_{1/2} \rightarrow 5d$) and L$_3$ ($2p_{3/2} \rightarrow 5d$) absorption edges. The energy of the incident x-ray beam was selected using a double-crystal diamond (1, 1, 1) monochromator. XAS measurements were performed at room temperature in transmission geometry to directly measure the linear x-ray attenuation coefficient, $\mu(E)$, using a series of three ion chambers. The sample was mounted between the first two, and IrO$_2$ powder was mounted between the last two ion chambers as a reference to ensure the energy calibration did not drift.

\subsection{NMR measurements} 
$^{19}$F NMR data were acquired at room temperature on a Bruker Advance 500 MHz (11.7 T) wide-bore NMR spectrometer and at a Larmor frequency of -470.6 MHz. The data were obtained under 50 kHz magic angle spinning (MAS) using a 1.3 mm double-resonance probe. $^{19}$F chemical shifts were referenced against lithium fluoride ($\delta$ = -204 ppm). The spin echo spectra were acquired on SrF$_2$ and Sr$_3$Ir$_2$O$_7$F$_x$ samples using a 90$^{\circ}$ RF pulse of 2.0 $\mu$s and a 180$^{\circ}$ RF pulse of 4.0 $\mu$s at 76.3 W. Recycle delays of 200 s and 12 s were used for SrF$_2$ and Sr$_3$Ir$_2$O$_7$F$_x$ samples, respectively. A $^{19}$F probe background spin echo spectrum, acquired under the same conditions as the spin echo spectrum of Sr$_3$Ir$_2$O$_7$F$_x$ samples but on an empty 1.3 mm rotor, revealed no significant background signal. Lineshape analysis was carried out using the SOLA lineshape simulation package within the Bruker Topspin software.

\subsection{DFT modeling} 
Theoretical results utilized density functional theory (DFT) within the {\it Vienna Ab-initio Simulation Package} (VASP)~\cite{VASP}, using the projector-augmented-wave method~\cite{PAW} and the Perdew-Burke-Ernzerhof (PBE) exchange-correlation functional~\cite{PBE}.  Spin-orbit coupling was included via the non-collinear spinor method implemented in VASP.  The DFT+$U$~\cite{Dudarev98} method was used to take into account correlation effects, and the $U$ value is chosen to be $1.6$ eV, based on constrained random-phase-approximation calculations from Ref.~\citenum{Kim17}.  This value was previously found to give a good description of the band structure and other electronic properties of Sr-327~\cite{Swift17}. 

\section{Results and discussion}

\subsection{Structural analysis}
Following reaction of Sr$_3$Ir$_2$O$_7$ crystals within a bed of CuF$_2$ powder, the reacted crystals were initially ground into a powder and checked with x-ray diffraction to explore whether the reaction spanned the bulk of the crystals.  Powder data collected from crushed crystals refined with a Lebail fit to a global structure in space group $C2/c$ with $a=24.2068(1)$, $b=5.45033(8)$, $c=5.50633(7)$, $\alpha=90$, $\beta=90.032(1)$, $\gamma=90$. This is consistent with the large $a$-axis expansion expected for the transformation from Sr$_3$Ir$_2$O$_7$ into Sr$_3$Ir$_2$O$_7$F$_2$ depicted in Fig. 1. Notably, the crushed crystals reacted under optimized conditions provide a powder with one unique $a$-axis lattice parameter, indicating a diffusion of F throughout the bulk of the crystals and with no unreacted Sr$_3$Ir$_2$O$_7$ remaining.   

To further parameterize the insertion of fluorine, quantitative $^{19}$F NMR spectra were collected on a SrF$_2$ standard and on a collection of crushed Sr$_3$Ir$_2$O$_7$F$_x$ single crystals to determine the F stoichiometry of the material. NMR spectra are plotted in Fig. 2 and reveal one major F environment with a chemical shift of $\delta=25.1$ ppm, consistent with the F ions inserted between the interstitial Sr atoms; however a weak secondary peak is also apparent between the centerband and first sideband of the primary F environment. The position of this minor F resonance matches with the chemical shift $\delta=88$ ppm expected for a SrF$_2$-like local environment \cite{schmidt2014sol}. For samples more aggressively fluorinated (longer exposure times), the small secondary resonance grows in relative intensity and reflects an eventual decomposition of the sample under prolonged exposure to fluorine.  

Under the optimized conditions reported here, the average fluorine content of the crushed crystals is measured to be Sr$_3$Ir$_2$O$_7$F$_{1.9}$ and the secondary SrF$_2$-like environment was too weak to be reliably quantified (ie. $<\approx 3\%$ molar fraction).  This small secondary peak could arise from either a slight overexposure to F$_2$ gas or from residual contamination of CuF$_2$ on the surface of one of the crystals crushed for analysis. Given this uncertainty and the inability to perform NMR measurements on the volume of a single small crystal, we will reference the compound as Sr$_3$Ir$_2$O$_7$F$_2$ for the purposes of this paper.  As will be shown later, this is supported by DFT calculations which demonstrate that the fully intercalated compound containing two F atoms per formula unit is the most stable, and any F deficiencies averaged across of crystal likely arise from phase separated Sr$_3$Ir$_2$O$_7$ regions. 

\begin{figure}[t]
	\includegraphics[scale=.15]{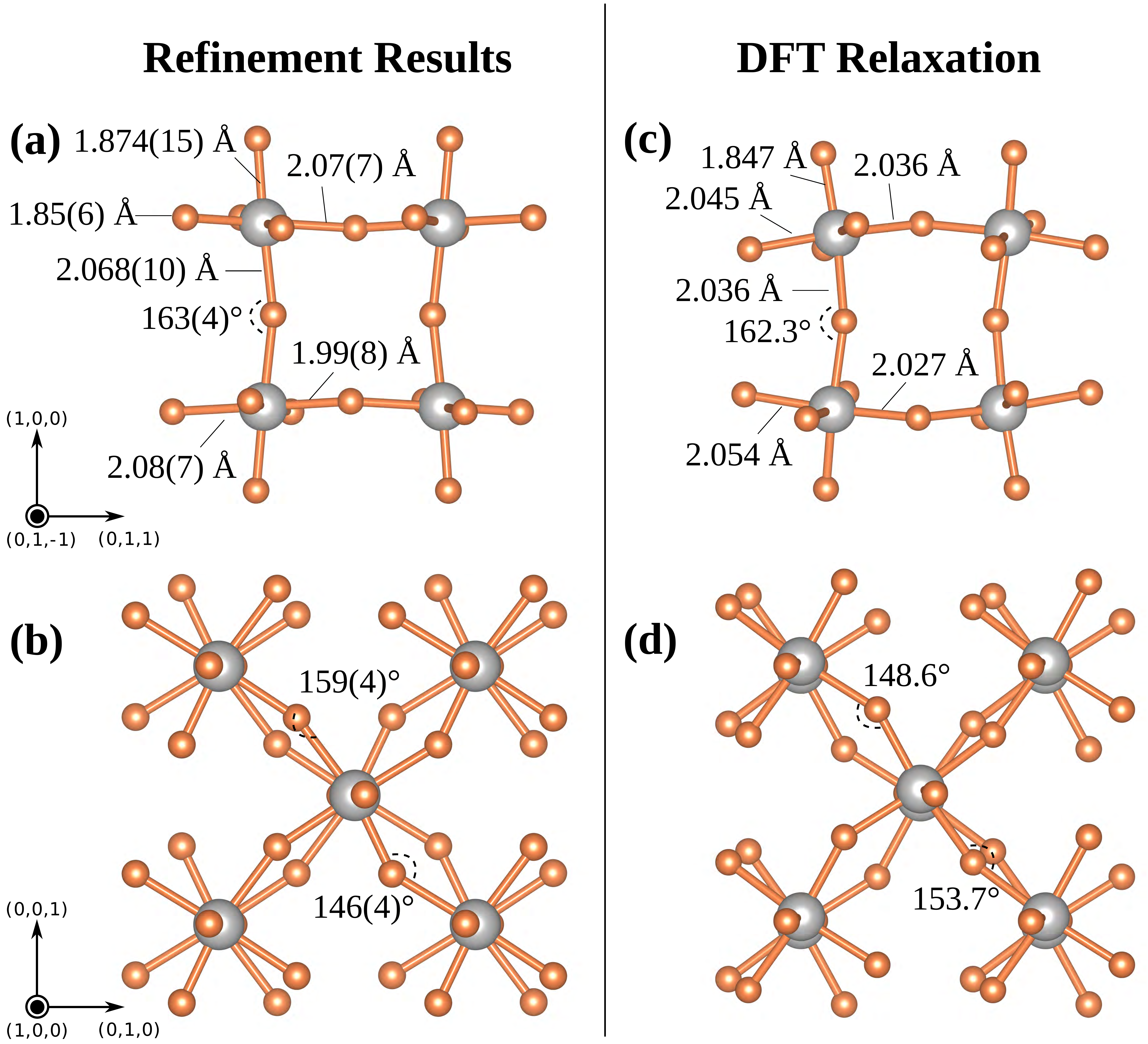}
	\caption{Illustration of the IrO$_6$ octehedra, Ir-O bondlengths, and Ir-O-Ir bond angles determined via (a),(b) x-ray/neutron diffraction and (c),(d) DFT structural relaxation calculations.}
\end{figure}

\begin{table}[t]
	\caption{Table of atom positions determined at 298 K using single crystal neutron and x-ray diffraction data within space group C2/c.  Isotropic thermal parameters are listed in units of $10^{-3}$ \AA$^{2}$. While isotropic thermal parameters were refined for x-ray data, neutron data were used to refine the atomic positions with constrained atomic displacements due to the limited accessible reciprocal space with 1.546 \AA.  These parameters were fixed at $3.8\times10^{-3}$ \AA$^{2}$ for Ir and Sr sites and $6.33\times10^{-3}$ \AA$^{2}$ for O and F sites for neutron refinement. Fit results yielded weighted R$_{F2}=0.074$, $\chi^2$=0.776 for x-ray data and weighted R$_{F2}=0.078$, $\chi^2$=3.07 for neutron data.}
	\label{tbl:example}
	\begin{ruledtabular}
	\begin{tabular}{ccllll}
		Atom  & site & x$_{xray}$ & y$_{xray}$ & z$_{xray}$ & U$_{iso}$\\
		\hline
		Ir & 8f & 0.58430(3) & 0.7653(8) & 0.7501(3) & 5.2(3)\\
		Sr1& 4e & 0.50000 & 0.225(2)&  0.75000&  10.1(12) \\
		Sr2& 8f & 0.68705(9) & 0.7790(15) & 0.2503(6)&  11.3(9) \\
		O1 & 4e & 0.50000 & 0.708(12) & 0.75000 & 16(11)\\
		O2 & 8f & 0.6611(6) & 0.803(6) & 0.754(4)& 5(6) \\
		O3 & 8f & 0.0890(13) & 0.410(13) & 0.447(10) & 31(11) \\
		O4 & 8f & 0.0794(11) & 0.962(14)&  0.551(10) & 23(8) \\
		F & 8f & 0.2558(12) & 0.505(8) & 0.003(12) & 16(4) \\
		\hline
		\hline
		Atom  & site & x$_{neutron}$ & y$_{neutron}$ & z$_{neutron}$ & U$_{iso}$ \\
		\hline
		Ir & 8f & 0.5872(8)  & 0.768(13) &  0.741(7) &3.8\\
		Sr1& 4e & 0.50000  & 0.210(15) & 0.7500 & 3.8\\
		Sr2& 8f & 0.6871(10) & 0.755(16) & 0.243(12) & 3.8\\
		O1 & 4e & 0.50000 & 0.72(2) & 0.75000 & 6.33\\
		O2 & 8f & 0.658(2) & 0.800(14) & 0.740(18) & 6.33\\
		O3 & 8f & 0.0699(15) & 0.488(13) & 0.533(16) & 6.33\\
		O4 & 8f & 0.088(2) & 0.933(13) & 0.589(15) & 6.33\\
		F & 8f & 0.252(8) & 0.509(9) & 0.00000 & 6.33\\		
	\end{tabular}
\end{ruledtabular}
\end{table}

A single Sr$_3$Ir$_2$O$_7$F$_2$ crystal was also measured with a combination of single crystal x-ray and neutron diffraction, and the structure was refined in the same twinned $C2/c$ unit cell as the parent Sr$_3$Ir$_2$O$_7$ material.\cite{Hogan16} As the crystal volumes are small and the mosaics of crystals are broadened following F insertion, the neutron data were collected with the high flux neutron wavelength (1.546 \AA) that limits the accessible reciprocal space for precisely determining the atomic displacement parameters. The neutron data was therefore analyzed in two ways:  (1) by applying a constraint fixing the Sr and Ir positions to those determined initially via the x-ray data and (2) by allowing all atomic positions to refine freely. Atomic positions determined via the two methods were consistent within error and the unconstrained refinement yielded better weighted R$_{F2}$ and $\chi^2$ fit parameters. While neutron diffraction is sensitive to the scattering from both O and F atoms, there is minimal scattering contrast in the coherent neutron cross sections between them.  Hence, F atoms were assumed to only occupy the same interstitial sites within the SrO planes as those identified in previous studies of Sr$_3$Ru$_2$O$_7$F$_2$ \cite{li2000double}; an assumption consistent with NMR data identifying a single F environment and the known expansion of the $a$-axis upon F insertion.  

Metal-oxygen and metal-fluorine bond lengths and bond angles determined via neutron diffraction agreed within error with the those of x-ray measurements, and the atomic positions are summarized in Table 1.  The x-ray refined structure is shown in Fig. 3 with the local oxygen octahedral environments and Ir-O-Ir bond angles are labeled.  The data reveal highly distorted IrO$_6$ octahedra with apical O sites closest to the F anions substantially compressed toward the Ir positions.  The spacing of Ir sites between the planes of a bilayer remains unchanged upon F insertion (4.0880(15) \AA~ for Sr$_3$Ir$_2$O$_7$F$_2$, 4.084(2) \AA~ for Sr$_3$Ir$_2$O$_7$), consistent with a model of F incorporated only in the planes between bilayers. 

A key difference between the parent material and the fluorinated compound is in the local iridium environments, where a strong deviation away from local cubic symmetry arises.  This can be parameterized via a distortion parameter for Sr$_3$Ir$_2$O$_7$F$_2$ of $\Delta_{d} = \frac{1}{6} \sum_{n=1,6}\big[ (d_n-d_{avg})/d_{avg} \big]^{2}=0.014$ where $d_n$ is summed over individual Ir-O bond lengths and $d_{avg}$ is the average Ir-O bond length.  Once F is inserted, this $\Delta_{d}$ becomes over an order of magnitude larger than the $\Delta_{d}=0.00066$ of the parent compound \cite{Hogan16} and presages a departure from the cubic limit.

The experimentally determined structure for Sr$_3$Ir$_2$O$_7$F$_2$ was further compared with a relaxed unit cell determined via DFT and plotted in Fig. 3.  The local IrO$_6$ octahedral environments in the theoretical and experimental structures agree reasonably well with one another, where the apical Ir-O2 bond length contracts and the Ir-O1-Ir interlayer bond angle is substantially distorted away from 180$^{\circ}$.  The expected positions of the F ions and the phasing of the octahedral distortions also mirror one another, and the modified long-axis lattice parameter $a_{exp}=24.24690$ \AA~ determined via single crystal x-ray measurements is close to the predicted value $a_{exp}=24.42624$ \AA. However, the DFT structure also predicts a larger orthorhombicity $b_{DFT}=5.46711$ \AA~ and $c_{DFT}=5.70697$ \AA~ than that observed experimentally $b_{exp}=5.45539$ \AA~ and $c_{exp}=5.48889$ \AA, and notable differences appear between the measured and calculated values for the in-plane Ir-O3 bond length. While twinning effects and disorder in the average structure of the crystal may account for some discrepancies, differences in the experimental $b$ and $c$ lattice parameters of the order predicted by DFT were not observed in either our x-ray or neutron diffraction measurements.     

\begin{figure}[t]
	\includegraphics[scale=.45]{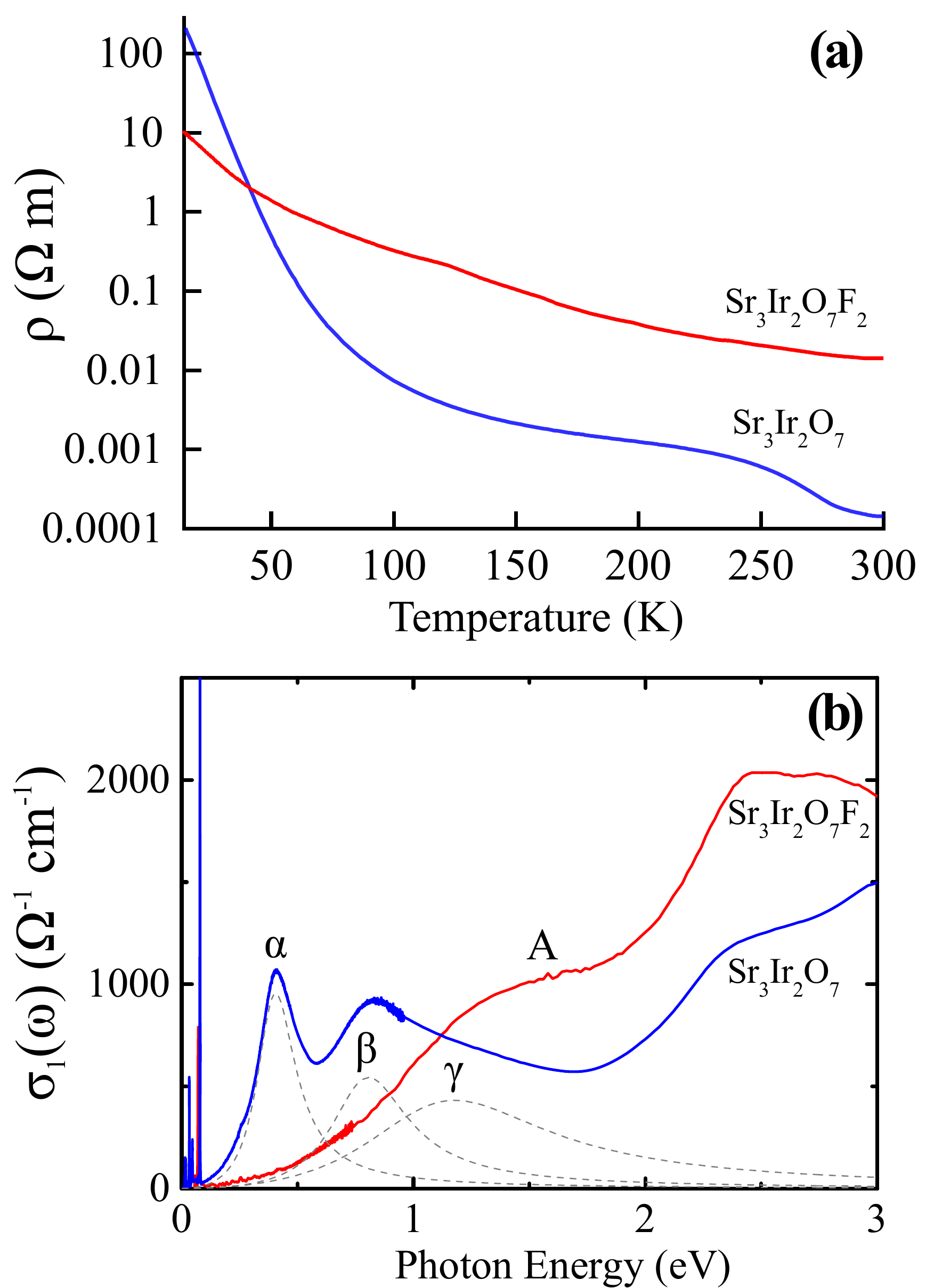}
	\caption{(a)  DC charge transport measured as a function of temperature for both Sr$_3$Ir$_2$O$_7$ and Sr$_3$Ir$_2$O$_7$F$_2$. (b) Optical conductivity measured at room temperature for both Sr$_3$Ir$_2$O$_7$ and Sr$_3$Ir$_2$O$_7$F$_2$.  $\alpha$, $\beta$, and $\gamma$ transitions are the expected transitions for the $J_{eff}=1/2$ Mott state of Sr$_3$Ir$_2$O$_7$ as described in the text, and the ``A" peak is labeled as the dominant excitation in Sr$_3$Ir$_2$O$_7$F$_2$.}
\end{figure}

\subsection{Electronic Properties of Sr$_3$Ir$_2$O$_7$F$_2$}
Electron transport and magnetization measurements were performed on single crystals of both Sr$_3$Ir$_2$O$_7$F$_2$ and Sr$_3$Ir$_2$O$_7$ in order to characterize the evolution of the electronic ground state as the iridium valence is shifted from Ir$^{4+}$ to Ir$^{5+}$.  Temperature dependent resistivity data with current flowing within the basal $bc$-plane are plotted in Fig. 4 (a).  The F intercalated system remains an insulator, and comparison with the resistivity of the crystal prior to fluorination shows that the room temperature resistivity increases by over two orders of magnitude. This implies an enhanced charge gap, and the weak divergence of Sr$_3$Ir$_2$O$_7$F$_2$'s resistivity upon cooling likely stems from in-gap impurity states due to disorder within the crystal introduced during the fluorination process.  Resistivity data did not fit to standard Arrhenius or variable-range hopping forms of transport.  

Exploring the gap structure further, optical conductivity measurements are plotted in Fig. 4 (b).  In Sr$_3$Ir$_2$O$_7$F$_2$, a charge gap appears with a slow build up of conductivity with a broad peak centered at 1.5 eV (labeled peak A) and a second, higher energy, feature at 2.5 eV.  In comparing the optical conductivity data with unfluorinated Sr$_3$Ir$_2$O$_7$, the lower energy transitions between the lower and upper Hubbard bands of the $J_{eff}=1/2$ manifold ($\alpha$) and the transitions from the $J_{eff}=3/2$ band into the upper Hubbard band ($\beta$ and $\gamma$) are suppressed. Instead, the F-intercalation seemingly drives the formation of a larger gap within the $t_{2g}$ manifold while the higher energy spectral features for $E>2$ eV are similar between the two compounds. 

Naively, in the single-ion limit, the formation of a $J=0$ state should have excited states with a first excited triplet $J=1$ band and a higher $J=2$ energy level.  The splitting of these states should be on the order of $\lambda\approx 0.2$ eV and $3\lambda \approx 0.6$ eV respectively, assuming a value for the spin-orbit coupling strength $\xi=0.4$ eV where $\lambda=\frac{\xi}{2S}$.\cite{kim2008novel} The band picture is considerably more complex, where the system's departure from the ideal limit of a cubic crystal field at Ir-sites as well as intermixing of oxygen states in the valence and conduction bands requires careful modeling. Furthermore, the optical conductivity spectrum of Sr$_3$Ir$_2$O$_7$F$_2$ revealing the single inherently broad ``A" peak is subject to selection rules of intersite hopping terms, which can potentially mask any lower energy bands.      

At a minimum, the optical data demonstrate the formation of a band insulating phase as the system transitions from an $5d^5$ into a $5d^4$ valence state. To test whether or not this insulating state is reflective of a $J=0$ ground state whose character is primarily driven via the strong spin-orbit coupling inherent to the Ir$^{5+}$ cations, x-ray absorption measurements (XAS) were performed.  XAS data collected at the Ir $L_3$ and $L_2$ absorption edges are plotted in Fig. 5 (a)  and illustrate an $L_3/L_2$ branching ratio of $6.2\pm 0.7$---an enhanced value consistent with the addition of holes and of strong spin-orbit effects in the electronic structure of this material \cite{PhysRevB.38.3158, clancy2012spin}.  Using the average number of six $5d$ holes, this value can be converted to a measurement of $<L\cdot S> = 3.5\pm0.3$ [$\mathrm{\hbar}^2$].\cite{PhysRevB.38.3158}  This value is again comparable to those observed in spin-orbit driven $J_{eff}=1/2$ Mott insulating iridates and suggests that the distortion of the IrO$_6$ octahedra in Sr$_3$Ir$_2$O$_7$F$_2$ is insufficient to appreciably broaden the bandwidth by mixing the spin-orbit split $t_{2g}$ bands (or alternatively by quenching the orbital moment of the $t_{2g}$ manifold entirely). As a result, the electronic ground state is consistent with the $J=0$ state expected from a low-spin $d^4$ valence in a cubic crystal field.    

\begin{figure}[t]
	\includegraphics[scale=.625]{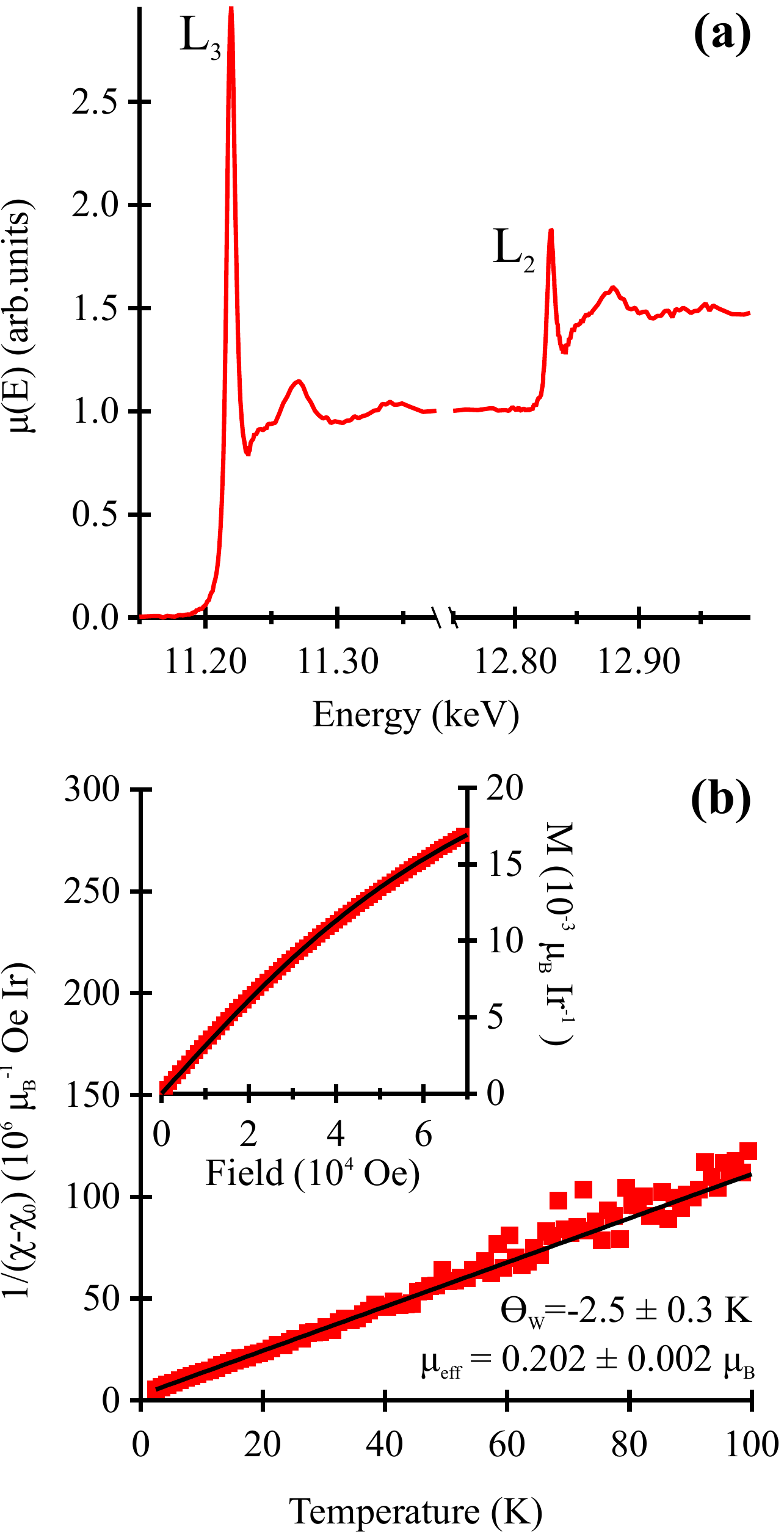}
	\caption{(a) X-ray absorption data showing the measured $\mu (E)$ at both the Ir L$_3$ and L$_2$ absorption edges. (b) Inverse susceptibility ($\chi-\chi_0$)$^{-1}$ plotted as a function of temperature. The temperature independent term in the susceptibility $\chi_0$ (comprised of the Van Vleck paramagnetic and background terms) has been removed to isolate the local moment response.  The resulting Curie-Weiss fit is shown as the black line through the data with values detailed in the text.  The inset shows field dependence of the magnetization collected at 2 K.  The solid line is a fit to a $J=1/2$ Brillouin function with a weak linear background as described in the text.}
\end{figure}

Magnetization measurements collected on a single crystal are plotted in Fig. 5 (b).  Naively, a $J=0$ ground state should exhibit only a weak Van Vleck paramagnetic response with quenched local moments. Our magnetization measurements are consistent with this expectation with the addition of a weak Curie-Weiss term that appears at low temperature.  The data were fit to the form $\chi(T)=\frac{C}{T-\Theta_W}+\chi_0$ with values $\chi_0 = 1.329\times10^{-7}$ $\frac{\mathrm{\mu_B}}{\textrm{Oe Ir}}$ and a Curie constant $C=9.227\times10^{-7}$ $\frac{\mathrm{\mu_B} \textrm{K}}{\textrm{Oe Ir}}$ yielding an effective moment of $0.203\pm0.002$ $\mu_B$ and Weiss temperature $\Theta_W=-2.5\pm0.3$ K.  This weak local moment response is commonly observed in Ir$^{5+}$ octahedrally coordinated compounds \cite{C5DT03188E,PhysRevB.91.155117} and often arises from a dilute concentration of magnetic impurities, such as local Ir$^{4+}$ $J_{eff}=1/2$ moments. This is  consistent with the majority of local moments being quenched within the $J=0$ state that forms in the low spin $d^4$ manifold.  

Assuming 1.73 $\mu_B$ per $J_{eff}=1/2$ impurity moment in the sample, the measured $\mu_{eff}$ would imply a $1.4\%$ concentration of remaining $J_{eff}=1/2$ moments.  To further verify this, the field dependence of the magnetization was measured at 2 K and the M(H) data are plotted in the inset of Fig. 5 (b).  These data are well fit by the Brillouin function expected for isolated $J_{eff}=1/2$ moments with a weak linear slope ($7.5\times10^{-4}$ $\frac{\mathrm{\mu_B}}{\textrm{kOe}}$) arising from the temperature independent Van Vleck and background contributions.  The saturation value for the Brillouin function fit is found to be $0.014\pm 0.004$ $\mu_B$ Ir$^{-1}$, in excellent agreement with the Curie-Weiss response expected to arise from a $1.4\%$ concentration of $J=1/2$ impurity moments.    

\subsection{\textit{Ab initio} modeling of Sr$_3$Ir$_2$O$_7$F$_2$}

The formation of the Sr$_3$Ir$_2$O$_7$F$_2$ structure was explored via calculations of partially-fluorinated structures in DFT calculations. Intermediate levels of fluorination were calculated by adding single fluorine atoms to the Sr-327 unit cell: from 0 (unfluorinated Sr$_3$Ir$_2$O$_7$) to 8 (fully fluorinated Sr$_3$Ir$_2$O$_7$F$_2$).  All symmetry-inequivalent sites were tested for each F atom added, and the structure was then fully relaxed (starting based on a linear interpolation of the two relaxed endpoints, Sr$_3$Ir$_2$O$_7$ and Sr$_3$Ir$_2$O$_7$F$_2$).  The fluorine site leading to the lowest energy after relaxation was chosen, and that site was kept occupied as further atoms were added. This allows for the absorption energy for $n$ fluorine atoms within the Sr$_3$Ir$_2$O$_7$ matrix to be determined as follows:
\begin{equation}
E_\text{Abs}(n) = E_\text{tot}[(\text{Sr-327})_4\text{F}_{n}] - 4E_\text{tot}[\text{Sr-327}] - n\mu_\text{F}\, .
\label{abs-eq}
\end{equation}

The chemical potential of fluorine (referenced to fluorine gas in the dilute limit) was then determined to be $\mu_F = - 0.797$ eV based on the vapor pressure of CuF$_2$ at 200 $^\circ$C,\cite{Brunetti08} assuming total dissociation into F$_2$ gas.  This also allows the calculation of the ``insertion energy'' of F, the energy change upon adding one additional fluorine atom to the unit cell:
\begin{equation}
E_\text{Ins}(n) = E_\text{Abs}(n) - E_\text{Abs}(n-1).
\label{ins-eq}
\end{equation}

These quantities are shown as a function of $x=n/4$, the number of fluorine atoms per formula unit, in Figure 6.  These results show that fluorine incorporation is strongly exothermic up to $x=2$, but moderately endothermic for higher $x$.  This agrees with the rapid, low temperature fluorination observed in experiments and explains why the system rapidly runs to complete fluorination $x=2$. Partial fluorination inevitably results in samples with regions of $x=2$ and $x=0$ concentrations, and potential annealing in the absence of a F source may allow for further staging behavior to occur. 

\begin{figure}
	\includegraphics[scale=.55]{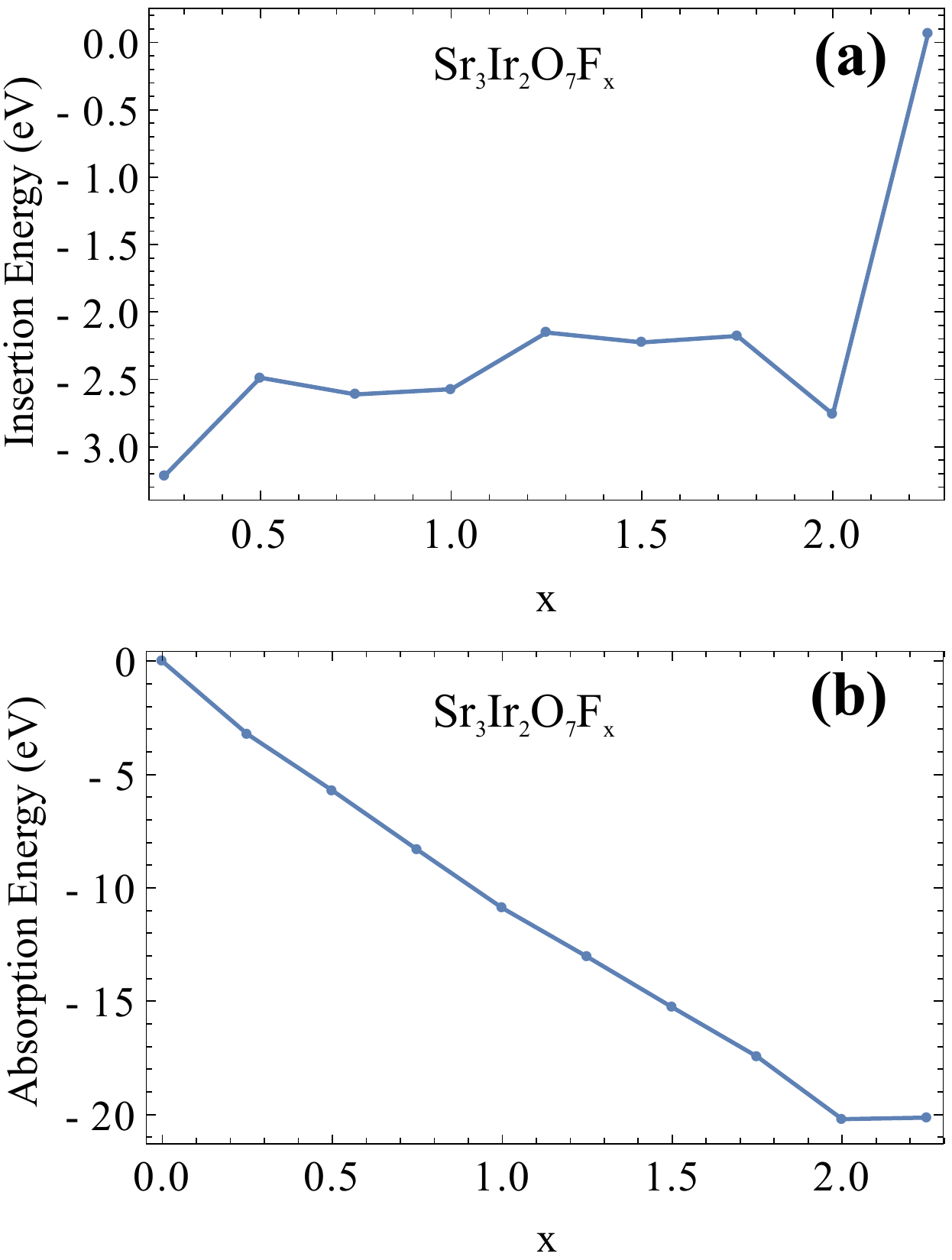}
	\caption{(a) Fluorine absorption energy (Eq.~\ref{abs-eq}) and (b) insertion energy (Eq.~\ref{ins-eq}) of Sr$_3$Ir$_2$O$_7$F$_x$.  This shows that fluorine incorporation will proceed spontaneously until $x=2$, in agreement with the synthesis results.}
\end{figure}

Electronic structure calculations were also performed to investigate the band structure of Sr$_3$Ir$_2$O$_7$F$_2$. The Brillouin zone was sampled using a $1\times4\times4$ $\Gamma$-centered grid---only a single {\bf k}-point point is needed in the $k_z$ direction because of the size of the long-axis lattice parameter and the quasi-2D nature of the material.  Using $U=1.6$ eV, the resulting band structure is plotted in Figure 7.  The 0.40 eV gap is indirect between the conduction-band minimum (CBM), located 19\% along X-$\Gamma$, and the valence-band maximum (VBM), located 50\% along X-$\Gamma$.  The VBM state has 53\% Ir $d$ character and 42\% O $p$ character, while the CBM has 61\% Ir $d$ character and 36\% O $p$ character.  If the $U$ term is removed from the calculation, the system remains gapped, confirming the band insulating character. Additionally, the valence band is comprised of comparable weights of $t_{2g}$ orbitals, consistent with the formation of a spin-orbit entangled $J=0$ valence state. However, the DFT model further predicts substantial spectral weight arising from $e_g$ orbitals within the valence band.  This does not conform to a simple picture of low spin $d^4$ electrons isolated in a spin-orbit quenched $t_{2g}$ multiplet, and instead suggests that the route to the $J=0$ ground state is more complex than a conventional $L\cdot S$-coupling scheme. 

Finally, the band structure was also computed in the absence of spin-orbit coupling (results not shown).  Without spin-orbit coupling there is no gap, and the Fermi level crosses Ir $d$ states in the minority spin channel.  These results reveal a band insulator whose charge gap is reliant on the strong spin-orbit coupling inherent to Ir cations in the system.

\begin{figure}
	\includegraphics[scale=.65]{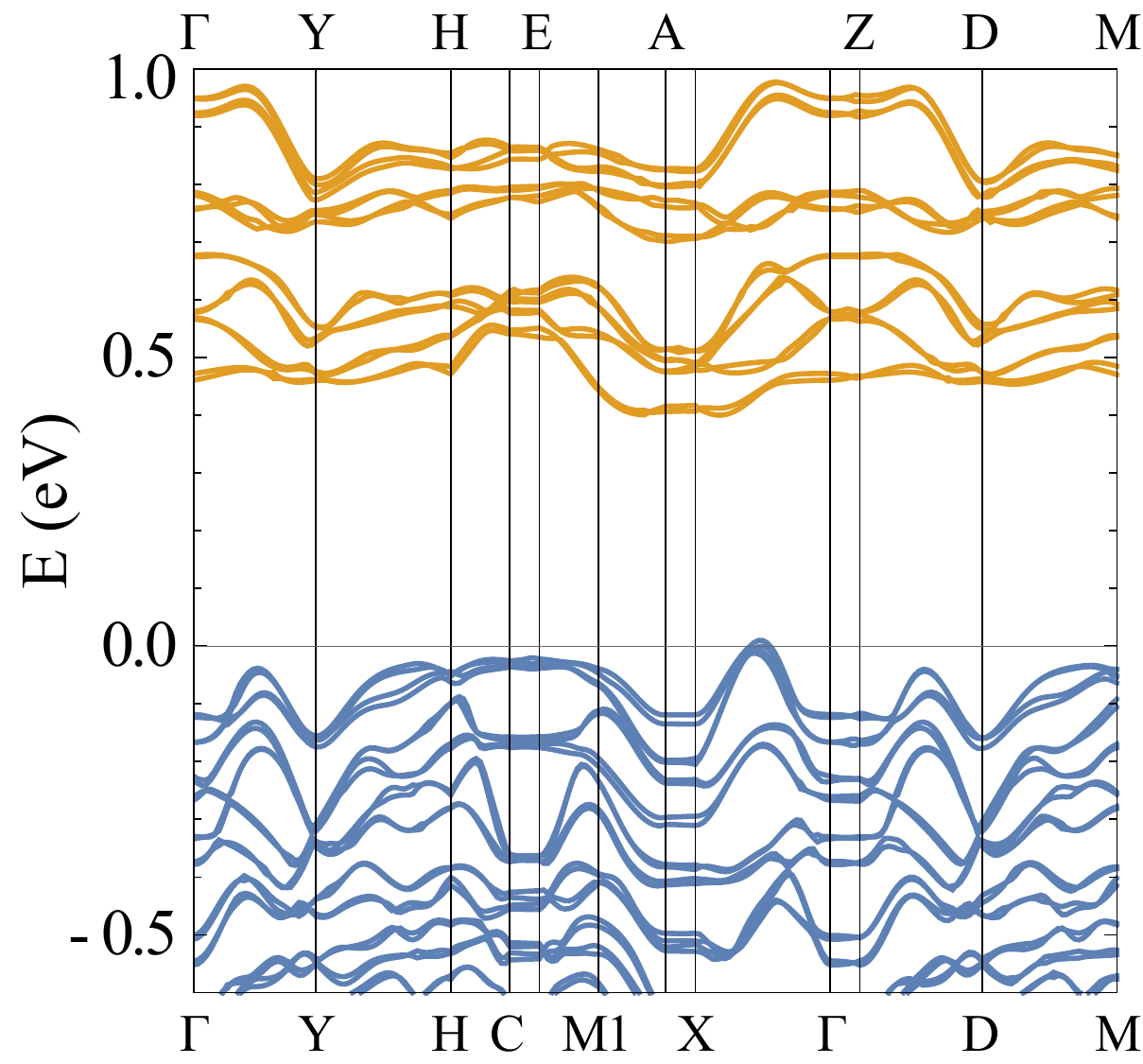}
	\caption{Band structure of Sr$_3$Ir$_2$O$_7$F$_2$ (cell defined such that the z-axis is aligned out of the iridium oxide planes).  Occupied valence bands are shown in blue, and unoccupied conduction bands are shown in orange.}
\end{figure}

\subsection{Discussion}

While topochemical transformations of R.P. phases via F insertion have been explored for a number of transition metal oxides, iridates are rather underexplored.  One obstacle has been the difficulty of preparing phase pure powder specimens of Sr$_3$Ir$_2$O$_7$---powder can only be stabilized under high pressure conditions \cite{nagai2007canted} and, even then, powder often includes Sr$_2$IrO$_4$ and SrIrO$_3$ impurities.  As a remedy, our results show that insertion of F anions can be directly realized within single crystals of Sr$_3$Ir$_2$O$_7$ where the reaction proceeds under relatively soft conditions (250 $^{\circ}$C for 2 hours).  

The insertion of F into Sr$_3$Ir$_2$O$_7$ parallels previous reactions of powders with dissociated CuF$_2$ in bilayer R.P. oxides such as La$_3$Ni$_2$O$_7$ \cite{zhang2016directed}, Sr$_3$Ru$_2$O$_7$ \cite{li2000double}, as well as in alloyed variants with mixed transition metal sites such as Sr$_3$(Mn$_{0.5}$Ru$_{0.5})_{2}$O$_{7}$ \cite{romero2013topochemical}. The shallow slope of the insertion energy in Sr$_3$Ir$_2$O$_7$ locally drives the complete transformation of Sr$_3$Ir$_2$O$_7$ into Sr$_3$Ir$_2$O$_7$F$_2$ such that crystals reacted for shorter periods of time are comprised of local patches of Sr$_3$Ir$_2$O$_7$F$_2$ and Sr$_3$Ir$_2$O$_7$.  Due to the large interface energy of these two phases within a single crystal, further staging behavior such as an intermediate Sr$_3$Ir$_2$O$_7$F phase may be realizable if the initial F loading and postgrowth annealing procedures are optimized. 

NMR measurements reveal that F insertion into Sr$_3$Ir$_2$O$_7$F$_2$ generates only a single local F environment, albeit one with substantial broadening. The main contribution to the NMR linewidth is likely due to residual homonuclear dipolar coupling interactions between the F spins in the layer, even under fast (50 kHz) magic angle spinning of the sample during data acquisition. Yet, we cannot exclude that local disorder also contributes to a lesser extent to the linewidth of the $^{19}$F resonance. 

The single resonant frequency attributed to Sr$_3$Ir$_2$O$_7$F$_2$ is consistent with F insertion only within the SrO rocksalt layers of the crystal matrix, and there is no indication of the partial anion exchange of F with apical O sites closest to F layers previously observed in Sr$_3$(Ti$_{0.5}$Ru$_{0.5}$)$_2$O$_6$F$_2$ \cite{romero2013topochemical} and Sr$_3$Fe$_{2}$O$_6$F$_2$\cite{A905730G}.  Scattering data further support this claim with the bond valence sums (BVS) \cite{brown2016chemical} of the nominal F site totaling 1.16 for x-ray data and 1.19 for the neutron data assuming F occupation versus 1.56 when assuming O occupation. The BVS of the nominal F sites assuming F occupation is also consistent other $n=2$ R.P. systems where F is known to insert between bilayers,\cite{zhang2016la2srcr2o7f2,li2000double} and the average Sr-F bond length of 2.47 $\AA$ is close to the 2.511 $\AA$ bond length of SrF$_2$ \cite{forsyth1989time}. 

Within the iridium-oxygen bilayers, the insertion of F affects an increased distortion of the IrO$_6$ octahedra and a commensurate increase in the bond valence sum of Ir cations from 4.18 in Sr$_3$Ir$_2$O$_7$ to 5.09 in Sr$_3$Ir$_2$O$_7$F$_2$.  This is consistent with the oxidation state of Ir being raised from Ir$^{4+}$ to Ir$^{5+}$ and the near complete intercalation of two F anions per formula unit.  The addition of F anions above and below the IrO$_6$ bilayers also drives a large contraction of the outermost apical Ir-O2 bond lengths from 2.022(2) $\AA$ in Sr$_3$Ir$_2$O$_7$ to 1.87(1) $\AA$ in Sr$_3$Ir$_2$O$_7$F$_2$.  This is consistent with other F intercalated bilayer systems that do not undergo anion exchange \cite{li2000double}.  While the overall extent of the bilayer blocks along the long-axis (distance between neighboring O2 sites along $a$) is shortened due to the compression of the Ir-O2 bond lengths, the distance between Ir atoms is unaffected due to the accompanying overall expansion of the $a$-axis.  

The change in Ir-O bond lengths and the increased distortion of the IrO$_6$ octahedra observed in scattering data are consistent with DFT calculations of the relaxed structure; however more precise comparisons of changes in the Ir-O-Ir bond angles upon F insertion are prohibited by the large uncertainties inherent to the single crystal neutron diffraction data.  This uncertainty is largely driven by a broadened crystal mosaic created by the F insertion process.  Upon F insertion, the crystal mosaic (defined by the full width at half maximum scattering intensity distribution of crystallites) increases from $<0.5^{\circ}$ to $\approx3^{\circ}$ in the larger (several mg) crystals suitable for neutron diffraction measurements.  DFT calculations show that the origin of the charge gap is predicated on the inclusion of spin-orbit coupling in the model, and spin-orbit coupling effects on the electronic ground state are further confirmed by XAS measurements that reveal a strongly enhanced branching ratio at the Ir L$_3$ and L$_2$ absorption edges.  This is consistent with the formation of a $J=0$ ground state within the low spin $5d^{4}$ valence electrons, despite the strong deviation of the local crystal field from perfect cubic symmetry. A number of other octahedrally coordinated  Ir$^{5+}$ systems have also been observed to manifest a $J=0$ ground state in lattice geometries varying from corner sharing double perovskites \cite{PhysRevB.93.014434} to edge-sharing honeycomb lattices \cite{C5DT03188E} to isolated octahedra.\cite{PhysRevB.91.155117} 

The lack of an appreciable local moment in the low field magnetization of Sr$_3$Ir$_2$O$_7$F$_2$ and, more specifically, the absence of fluctuation-driven magnetic order is notable. There is considerable interest in investigating the possibility of a spin-orbital liquid state \cite{PhysRevLett.116.097205, PhysRevLett.111.197201} as well as other unconventional magnetic states \cite{PhysRevB.91.054412} in strongly spin-orbit coupled $d^4$ $J=0$ spin systems. The conversion of Sr$_3$Ir$_2$O$_7$, a system thought inherently close to a dimer instability \cite{PhysRevB.92.024405, PhysRevB.94.100401}, into Sr$_3$Ir$_2$O$_7$F$_2$ is a test case for exploring the presence of these electronic phases. Sr$_3$Ir$_2$O$_7$ itself also lacks a high temperature Curie-Weiss response above its antiferromagnetic ordering temperature, and any rare, phase separated regions of Sr$_3$Ir$_2$O$_7$ within the crystal naively should not yield a measurable local moment.  Instead the measured $\mu_{eff}\approx 0.2$ $\mu_B$ reflects a dilute concentration of impurity moments likely arising from another defect mechanism such as a low concentration of oxygen vacancies.   

\section{Conclusions}

We have demonstrated that single crystals of the $J_{eff}=1/2$ Mott insulator Sr$_3$Ir$_2$O$_7$ can be converted into the $J=0$ band insulator Sr$_3$Ir$_2$O$_7$F$_2$ via the insertion of fluorine into the crystal matrix.  Reaction of crystals embedded within CuF$_2$ powder proceeds rapidly and converts the entirety of Ir$^{4+}$ cations into Ir$^{5+}$ with negligible anion exchange observed. F atoms insert within the SrO interstitial planes between the bilayers of the lattice, resulting in a strong perturbation to the IrO$_6$ octahedral environments away from an ideal cubic symmetry.  Despite this distortion, x-ray absorption measurements and first principles calculations are consistent with the formation of a spin-orbit entangled $J=0$ electronic ground state that possesses a charge gap reliant on the strong spin-orbit coupling inherent to the Ir cations.

\section{Acknowledgments}
This work was supported primarily by ARO Award W911NF-16-1-0361 (S.D.W., C.P., Z. P.). M. W. S. was supported by the MRSEC Program of the National Science Foundation under Award No. DMR-1121053.  The MRL Shared Experimental Facilities are supported by the MRSEC Program of the NSF under Award No. DMR 1720256; a member of the NSF-funded Materials Research Facilities Network.  The work at HYU was supported by Basic Science Research Program through the National Research Foundation of Korea (NRF) funded by the Ministry of Science, ICT and Future Planning (Grant No. 2017R1A2B4009413).  The work at ORNL's HFIR was sponsored by the Scientific User Facilities Division, Office of Science, Basic Energy Sciences, U.S. Department of Energy. Research conducted at CHESS is supported by the NSF under award DMR-1332208.

\bibliography{BibTex_Sr3Ir2O7F2}

\end{document}